# From "Analogue" Science to AI-powered Digital Science

## 1. Introduction

Historically, science has been advanced by the brightest minds collecting the cutting-edge information available at their times, analyzing it and then synthesizing it into knowledge, bringing together the general picture of how the universe works. This approach has been working well for centuries, when the amount of information that needed to be processed by a person's brain was relatively small. However, old methods are letting us down now that the information volume sky rockets. The amount of information produced daily goes far beyond the capacity of an individual scientist to digest it even within one discipline, let alone to bring together scattered pieces of knowledge from different fields to create a harmonized model of nature. The processing power of a biological brain is no longer sufficient to synthesize the overwhelming flow of research information. The information yield is decreasing, and, if I can frivolously borrow the term from the mathematical optimization theory, the "information to knowledge" loss function is steadily increasing.

Widespread use of computer algorithms for facilitating the scientific investigation could help us circumvent this problem. Nowadays, machines are used routinely for data acquisition and analysis, but not so much for data synthesis. Whether it is because of the vanity or just too recent availability of suitable computer technologies, but we people still keep the higher-order cognitive functions of information integration largely for ourselves. Despite the fact that it's only computers that could potentially integrate all the available research data and help humans make sense of it, most of the knowledge is still produced in the form of manuscript publications, which doesn't allow its apprehension by computers.

Nevertheless, the phase transition from the human-limited, "analogue" way of research enquiry to the silicon-based, artificial intelligence (AI)-powered digital science is forthcoming. Paraphrasing the words of [literally] rocket science pioneer K. Tsiolkovsky, "the biological brain is the cradle of science, but science cannot stay in the cradle forever". To facilitate this transition, I have come up with three project ideas: 1) *Crypto*Science platform, aimed to provide tools for endemic digitalization & open access of the research data, with attribution of the work credit of all involved individuals and parties using blockchain technology; 2) Computational Publication Standard, designed for publishing research findings in an AI-friendly format of knowledge graphs, that will render them easily readable by machines and available for seamless integration into mathematical models; 3) data modelling from publications, meant to extract key pieces of knowledge (entities and their relationships) contained in manuscript publications using natural language processing and image recognition, to enable their incorporation into the digital knowledge databases. These three ideas could be implemented separately by different research groups and institutions, however, they will strongly benefit from each other's synergistic effect, in my opinion.

## 2.1. *Crypto*Science: Need

Currently, science ecosystem is facing many systemic problems. The system of journal publications resembles the theatre of the absurd: every year, scientists pay thousands of euros for their findings to be published, institutions pay millions of euros for the researchers' journal subscriptions, and for an average person, scientific knowledge is still behind the paywall. Data reuse, sharing and open science are largely not adopted by the general ~~public~~ scientific community. (Un)natural selection of data is common, when experimental data that "fit the story" get cherry-picked and published, while other pieces of evidence – especially those that contradict the investigator's hypothesis – are excluded, which worsens the looming replication crisis. Moreover, people are naturally biased, and often see everything from the prospective of their research focus, while ignoring facts and factors beyond their comfort zone. This may not be a problem as such, but awareness and reflection upon one's biases should be important. Dogmatically established theories and opinions are hard to challenge, and early career researchers, who are more likely to come up with fresh ideas, are often unheard, until they mature into established scientists – who at this point become themselves reluctant to challenge the status quo. New research ideas and data are kept secret from potential "peer competitors" for years, before being "triumphantly" revealed in a publication, further stagnating the progress. Finally, objective measure of one's skills and performance is missing, and assessments are made based on publications and references, which is quite an indirect and hardly adequate way.

### 2.2. *Crypto*Science: Approach

*Crypto*Science is a web-based platform for universal digital representation of scientific data in both human- and AI-friendly format. The key concept behind *Crypto*Science is utilization of blockchain technology that will enable any operations with scientific information and data to be metatagged with the contributor information. This will allow for immediate recognition and documentation of one's intellectual contribution, removing multiple conventional barriers and paving the way for open and transparent research process.

Specifically, *Crypto*Science aims at delivering the following functionality:

- Encrypted processes of data annotation, organization, storage, analysis, reanalysis, visualization, sharing and public access, when all operations with the data will be recorded as metadata using blockchain technology;
- Verification and attribution of one's own intellectual property, including scientific ideas and hypotheses;
- Personal track record of a user's activity, experience, expertise and contributions to the field;
- Publication of experimental results with direct in-built reference to the experimental data, protocols and other resources, including literature;
- Facilitation of peer communication, evaluation and help; peer review timely available at the stage of planning, running & analyzing the experiments, when it is naturally appropriate and needed, not only at the final post-production stage, when it is often tardy and psychologically unwanted;
- Establishment of scientific collaborations based on one's scientific interests and expertise, circumventing traditional hierarchies and other barriers.

### 2.3. *Crypto*Science: Benefits

- *Crypto*Science automatically verifies the user's contribution when publishing experimental data, ideas, hypotheses and other intellectual contributions to the field. -> No need to hide scientific data & ideas from the community and the public for years. Access restrictions, sensitive metadata anonymization and time lags for publication can be introduced, if necessary.
- Easy access to data, tools and expertise for everyone.
- "Representativity" of the data – all experimental data go there, not just those handpicked for publication.
- Transparency and quality assurance of the data.

- Easy and open publication process, making journals redundant, and publication fees obsolete.
- Facilitated and open peer review at all stages of the research.
- Easy cross-linking of the papers, protocols, data sheets etc.
- Common intellectual and data space to facilitate exchange and development of ideas.
- Personal track record: theoretical and experimental impact & significance to the field of an individual, influence in the community etc.
- Early career researchers having their say (ideas, analysis of previously published data etc.)
- Community-developed experimental methods, analysis and visualization tools.
- Self-evaluation of biases in terms of publications and other resources that have been read and used for references.

### 2.4. *Crypto*Science: Commercial potential

One global impact that *Crypto*Science platform can have is being a tool for commercialization of knowledge. *Crypto*Science may be thought of as an intellectual property (IP) marketplace, where one's ideas, knowledge or research findings can be monetized directly. While data (re)use for research should always be free, it can bring profit to the original contributor(s) if their findings are used for commercial purposes.

*E.g.: Currently, pharma companies pay millions of euros for running or outsourcing FDA-required tests. A lion's share of these expenses end up being wasted, since the ideas for (pre)clinical trials are based on academic research that is often carried out with major violations of quality assurance. However, if companies could directly use academic research findings in their portfolios for FDA applications and recompense the contributor researchers for those,*

*it would benefit: the companies themselves, in terms of time- and cost-savings; the academic labs, in terms of money compensation; the science itself, because scientists will see incentives in adopting good laboratory practice (GLP) and the highest research standards in their research pipelines.*

### 3.1. Computational Publication Standard: Need

The format of manuscript publication in journals is largely a conventional heritage of the past when the articles had to be printed on paper and physically distributed to the target audience. However, carrying things over this way nowadays severely limits the efficiency of data usage in several ways.

Firstly, it takes time even for an experienced reader to extract the most important information from a paper and to compare it with other published sources. In case conflicting evidence exists, it's quite a tedious process to look for the experimental details possibly underlying the differences in experimental outcomes.

*E.g.: When working on my PhD thesis, I have read a lot about the effect of perineuronal net (PNN) formations on the physiology of parvalbumin-positive (PV+) neurons. Some authors found that PNN removal generally promotes PV+ cell activity* (Hayani et al., 2018; Shi et al., 2019), *while others came up with completely opposite findings* (Balmer, 2016; Lensjø et al., 2017; Tewari et al., 2018). *Investigating the reasons behind these experimental discrepancies was time-consuming, and I'm still lacking a decisive conclusion about this topic.*

Secondly, and most importantly, a manuscript publication is a format that is not easily accessible and operandable by machines. Publishing raw data in depositories is a good step towards enabling further computational modelling. However, there is still a number of issues that hamper the data utilization for modelling, such as incompatibility of formats and poor quality of data annotation. Finally, dimensionality reduction algorithms is an important problem that still stays largely unresolved.

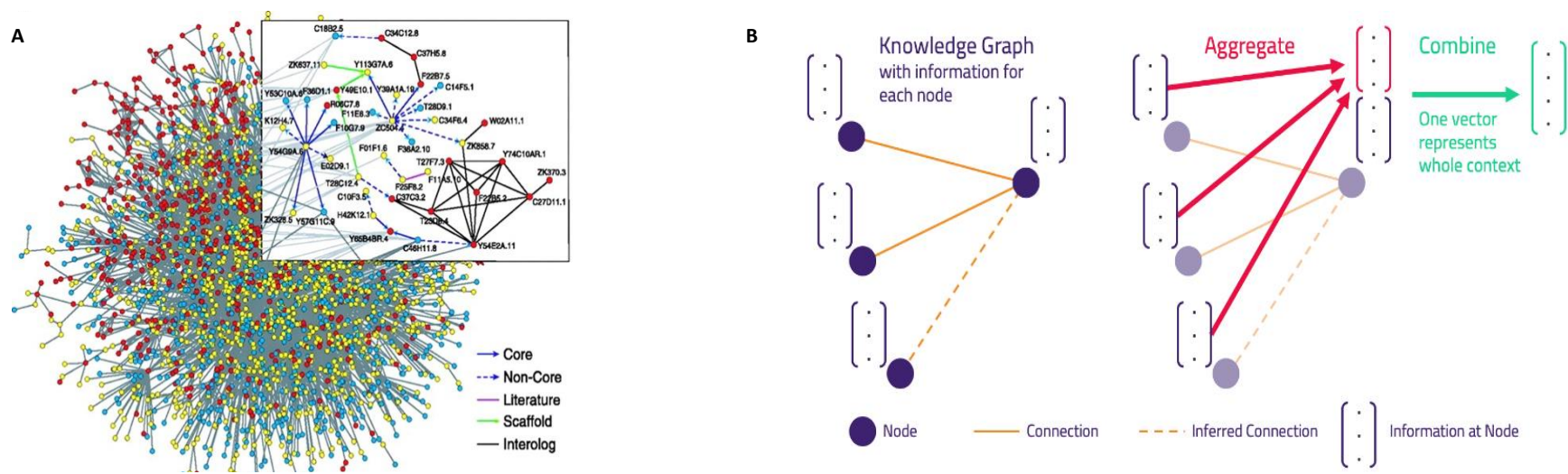

Figure 1. (A) Example demonstrating visual knowledge graph representation with graph zoom-in (From Li et al., 2004. Reprinted with permission from AAAS). (B) Example demonstrating graph node connection inference and information (including metadata) attachment (diagram based off of Vaticle's open-source KGCN project. Reproduced with permission from Vaticle).

## 3.2. Computational Publication Standard: Approach

The idea behind introducing Computational Publication Standard is to represent research findings in the form of knowledge graphs (Knowledge graphs|The Alan Turing Institute) that will describe research entities and relationships between them. Importantly, the graph nodes (and edges) will be a) multidimensional, allowing for representation and testing of rival hypotheses, b) nested, allowing for zooming between different node levels to enable enclosure of sub- or supra-graphs, c) hybrid, allowing for node/edge type description, cross-linking with raw datasets and metadata attachment. Introduction of Computational Publication Standard in science could be compared to the invention of Lego bricks, as it allows for seamless integration of pieces of knowledge within a single mathematical framework.

## 3.3. Computational Publication Standard: Benefits

- Experimental data are published in an AI-friendly format, rendering their analysis and computational modelling malleable to be run by machines;
- Graphs could be created based on the data of your choice, allowing for comparing separately chosen experimental results and testing hypotheses with different initial variables;
- Inference engineering of hidden connections between the entities that haven't been previously revealed could be made;
- Ideally, compatibility between different levels of organization could be achieved, so that the information acquired at lower levels (e.g. molecular structures) could be used, perhaps with less detailization, in higher-order models (e.g. when neuronal firing is simulated).

### 4.1. Data Modelling from Publications: Need

Computational Publication Standard could significantly formalize our scientific knowledge, rendering it immediately available for mathematical modelling and computer simulations. However, there is a large corpus of experimental data that has already been produced by the humanity in the last 100 years but that cannot currently be interpreted and used by computers.

### 4.2. Data Modelling from Publications: Approach

This knowledge could potentially be formalized and expressed in graphs using natural language processing (NLP) and pattern recognition techniques (to analyze figures and bars) that are quickly progressing thanks to recent advances in machine learning (ML) and AI. The "resolution" and content saturation of such knowledge graphs will be inherently lower than that of graphs originally created within Computational Publication Standard ecosphere, but will still allow for key entity and relationship extraction. Information extracted using NLP and figure analysis could be compared and self-validated/self-corrected, improving the accuracy and efficiency of both techniques.

*E.g.: Current ML algorithms should be able to extract pieces of knowledge such as "Drug X increases expression of protein Y in type Z of neurons by 30%", and create a graph based on those.*

### 4.3. Data Modelling from Publications: Benefits

- Salvaging of historical scientific data for their adaptation in the formal algorithms that will be employed by science of the future.

E-mail: angelina.lesnikova@helsinki.fi
Twitter: @an_lesnikova